\documentclass[onecolumn,manuscript,doublespace]{svjour2}
\usepackage{graphicx}

\begin{document}

\title{Partial measurements and the realization of quantum-mechanical counterfactuals}

\author{G.~S.~Paraoanu}
\institute{G.~S.~Paraoanu \at
Low Temperature Laboratory, Aalto University, P. O. Box 15100, FI-00076 AALTO, Finland \email{paraoanu@cc.hut.fi}
\at
Institute for Quantum Optics and Quantum Information (IQOQI),
Austrian Academy of Sciences, Boltzmanngasse 3, A-1090 Vienna, Austria}


\maketitle

\begin{abstract}

We propose partial measurements as a conceptual tool to understand how to operate with counterfactual claims in quantum physics. Indeed, unlike standard von Neumann measurements, partial measurements can be reversed probabilistically. We first analyze the consequences of this rather unusual feature for the principle of superposition, for the complementarity principle, and for the issue of hidden variables. Then we move on to exploring non-local contexts, by reformulating the EPR paradox, the quantum teleportation experiment, and the entanglement-swapping protocol
for the situation in which one uses partial measurements followed by their stochastic reversal. This leads to a number of counter-intuitive results, which are shown to be resolved if we give up the the idea of attributing reality to the wavefunction of a single quantum system.
\end{abstract}


\section{Introduction}

A bipartite system, prepared in an entangled state, has correlations stronger than those resulting from the use of local, classical variables \cite{peresbook}. In the modern theory of quantum information,
this peculiar feature is exploited much like a resource: for example quantum computers would make use of these correlations to speed up certain tasks, thus indicating that encoding  every bit of information in a classical variable is not always an efficient way of performing computations.

At the same time, the non-separability of entangled states has been at the core of fascinating debates since the beginning of quantum physics. Part of the origin of the well-known quantum-mechanical "paradoxes" is the degree to which "reality" and "information" become mixed and engrained in the quantum-mechanical wavefunction. This clearly shows up in situations such as EPR, teleportation and entanglement swapping, where simply acquiring information from a distant, spatially-separated source seems to change the physical quantum-mechanical description of  the system in study, prompting Einstein to conclude that this description is therefore necessarily incomplete \cite{epr}. One solution out of this conundrum is to take {\it ad litteram} the idea that quantum physics is all about the information we can acquire about a physical system in a classically well-specified setup. This view can be regarded as a modern-day Copenhagen interpretation. It emphasizes that the origin of our difficulties in understanding quantum mechanics is the tendency to reason counterfactually. But in quantum mechanics statements like "what would have been the result if we have had measured the x-component of  the spin instead of the y-component" on a given single quantum system simply do not make any sense \cite{peres}. A good {\it dictum} for this situation is "unperformed experiments
have no results" \cite{peresamjphys}.
In Bohr's view, once the classical measuring arrangement for an observable is in place, this defines a configuration of  the Universe which is incompatible with the configuration for measuring a conjugate observable. For example, in the case of spin -1/2, the configuration for a measurement along the $x$ direction defines the resulting state of the quantum
after the measurement (spin oriented along $x$), and it does not make sense to try to infer, after the measurement is performed, what would the state have been had it been measured along $z$.

In this paper we attempt to see if it is possible (and if so, how) to make counterfactual reasonings by involving partial measurements instead of the standard von Neumann measurements. The von Neumann measurements are also referred to as  "projective measurements" or "sharp measurements". Partial measurements are fully compatible with the framework of quantum mechanics: they can be regarded as generalizations of von Neumann measurements, and they are described by more general measurement operators which are not necessarily projectors.  Partial measurements, as we shall see, belong to the class of POVM measurements \cite{peresbook,nielsen,povm}, which are also called "unsharp measurements". Partial measurements are nowadays available in the lab. They have been developed for phase qubits \cite{partialmartinis}, as well as for charge-phase qubits in the context of interaction-free experiments \cite{interactionfree}. In the latter context, these measurements can be used to perform tasks with no classical analog, such as the detection of a pulse current without any energy absorption. Experimental implementations using photonic qubits are also possible \cite{kim}.

While the standard measurement process in quantum physics is irreversible, partial measurements have the interesting property that they can be undone (reversed) in a probabilistic way. Thus, although the associated measurement operators are nonunitary, an inverse still exists \cite{korotkovjordan}, and therefore partial measurements share with unitary evolution the feature of reversibility. As we will see, it is possible in some sense to "undo" the measurement conditionally, and reverse the qubit back to its initial state \cite{katz}.

In this work, the consequences of this stochastic reversibility are examined in the context of standard experiments testing the foundations of quantum physics. We argue that the idea that the wavefunction is a representation of a real entity is untenable. The paper is organized as follows: we introduce partial measurements and derive some of their properties in Section \ref{part}. In the related Appendix \ref{ap1} a generalized version of partial measurements is presented. In Section \ref{sup} we then show how entanglement is a consequence of systematic application of the superposition principle. The connection with hidden variables is examined in Section \ref{hid} and Appendix \ref{ap2}, and that with the principle of complementarity in Section \ref{comp}. In Section \ref{epr} we look at the EPR paradox with partial measurements, and we further explore the consequences for quantum teleportation (Section \ref{tel}) and entanglement swapping (Section \ref{swap}). The paper ends with conclusions (Section \ref{conc}).

\section{Doing and undoing partial measurements}
\label{part}

\paragraph{Partial measurements: generalities}

Partial measurements can appear in many contexts in quantum physics. As mentioned before, they are a particular type of the more general class of POVM measurements which generalize the standard von Neumann projective measurements \cite{povm}.
For a single qubit (with states $|0\rangle$ and $|1\rangle$)  a single parameter --  the so-called partial measurement strength $p$ ($0\leq p\leq 1$) -- is used to define two measurement operators $M_{\bar{m}}$ and $M_{m}$ (with measurement results $\bar{m}$ and $m$), by
\begin{eqnarray}
M_{\bar{m}} &=& \sqrt{p}|1\rangle\langle 1| ,\label{em}\\
M_{m} &=& |0\rangle\langle 0| + \sqrt{1-p}|1\rangle\langle 1| .\label{embar}
\end{eqnarray}
Note that, unlike the case of standard von Neumann measurements, the
operators $M_{\bar{m}}$ and $M_{m}$ are not necessarily projectors, and also that $M_{\bar{m}}M_{m}\neq 0$ if $p\neq 1$; projective measurements are obtained if and only if $p=1$. However, the results $\bar{m}$ and $m$ do mean that either one of them occurs (the bar above $m$ signifies logical negation, the "not obtaining $m$" in an experiment). Also, there is no stringent requirement of why the operator $M_{\bar{m}}$ should be constructed from only one projector: in fact it can have the same structure as $M_{m}$. This generalization is given in Appendix \ref{ap1}.
The POVM elements (also called "effects") associated with the measurement are positive operators defined (with standard notations) as
\begin{eqnarray}
E_{\bar{m}} &=& M_{\bar{m}}^{\dag}M_{\bar{m}} = p |1\rangle\langle 1|  , \\
E_{m} &=& M_{m}^{\dag}M_{m} = |0\rangle\langle 0| + (1-p) |1\rangle\langle 1| .
\end{eqnarray}
The operators $E_{\bar{m}}$ and $E_{m}$ satisfy the relation $E_{\bar{m}}+E_{m}=1$, which is called a semispectral resolution of identity (the standard situation, in which the $E$'s are projectors, is referred to as spectral resolution).
The effects $E_{m}$ and $E_{\bar{m}}$ can be used to define, given an initial pure state $|\psi\rangle$, the respective (conditional) probabilities $P(\bar{m}|\psi)$, $P(m |\psi)$ (with $P(\bar{m}|\psi)+P(m|\psi)=1$) of the outcomes $\bar{m}$ and $m$ in the standard way \cite{nielsen}
\begin{eqnarray}
P(\bar{m}|\psi) &=& \langle \psi |E_{\bar{m}}|\psi\rangle ,\label{uno}\\
P(m|\psi) &=& \langle \psi |E_{m}|\psi\rangle .\label{duo}
\end{eqnarray}
The corresponding wavefunctions resulting after the measurement depend on which result, $\bar{m}$ or $m$, has been obtained,
\begin{eqnarray}
|\psi_{\bar{m}}\rangle &=& \frac{1}{\sqrt{P(\bar{m}|\psi)}}M_{\bar{m}}|\psi\rangle ,\label{aftershave}\\
|\psi_{m}\rangle &=& \frac{1}{\sqrt{P(m|\psi)}}M_{m}|\psi\rangle .\label{after}
\end{eqnarray}
\begin{figure}[t]
\begin{center}
  \includegraphics[width=12cm]{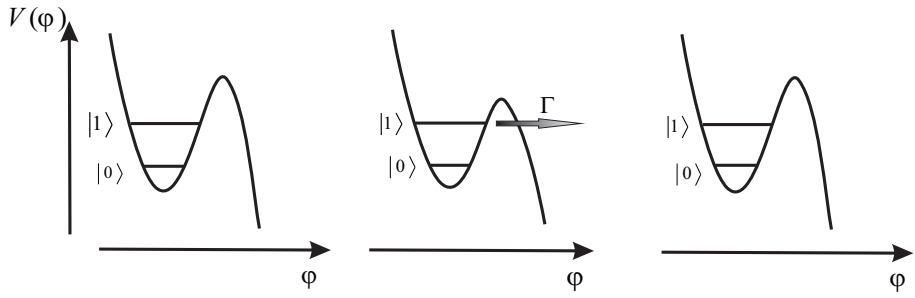}
\end{center}
\caption{Schematic of the qubit manipulation during a partial measurement. The
potential $V(\varphi)$ experienced by the qubit is changed such that, during a time $\tau$, the qubit can tunnel out if it is in the state $|1\rangle$, with tunneling rate $\Gamma$ (middle figure).}
         \label{fig1}
\end{figure}

\paragraph{Interpretation as a measurement on an ancilla: the case of superconducting qubits}

We now turn to describing how the formalism outlined above is implemented in the
the experiments with superconducting qubits.  Here $\bar{m}$ corresponds to a current-switching event, during which the wavefunction of the system -- which is initially localized in one of the wells of a Josephson washboard potential -- tunnels out in the quasi-continuum \cite{me}, resulting in the occurrence of a non-zero voltage across the junction. In contrast, $m$ corresponds to the recording of a zero-voltage, indicating that tunneling did not occur. Using the simple model for tunneling in quasi-continuum described in \cite{me}, we want now to show that partial measurements
can be regarded as a combination of unitary evolution on an enlarge Hilbert space containing an ancilla, followed by a projective measurement on the ancilla, a result which is generic for all POVMs \cite{nielsen}. In this model, the Josephson junction is biased for some time $\tau$ at a bias current chosen such that
the tunneling rate for the state $|0\rangle$ is negligible, while for the state $|1\rangle$ it has a finite value $\Gamma$ yielding a tunneling probability in the interval $\tau$ of $p = 1-\exp(-\Gamma \tau )$ (see Fig. \ref{fig1}). Also, the changes in the potential $V(\varphi )$ ($\varphi$ is the macroscopic superconducting phase across the junction) are adiabatic (slow) with respect to the timescale given by the inverse of the energy separation between the levels $|0\rangle$ and $|1\rangle$, but they are instantaneous when compared to $\Gamma^{-1}$.

We leave aside the qubit's Hamiltonian evolution in the interval $\tau$ (which is usually either negligible or irrelevant for our arguments below).  Suppose now that the qubit is in state $|1\rangle$, which is the first excited state $\psi_{1}(\varphi ) = \langle \varphi  |1\rangle$, localized in one of the wells ($\varphi$ is the phase variable). The evolution of this state during the measurement is \cite{me}
\begin{equation}
\psi_{1}(\varphi ) \rightarrow
\exp (-\Gamma \tau /2) \psi_{1}(\varphi ) + \psi_{1}^{\rm (out)} (\varphi )  ,\label{pp}
\end{equation}
where $\psi_{1}^{\rm (out)} (\varphi )$
corresponds to a propagating wavepacket that can be calculated explicitly and which results in the appearance of a macroscopic voltage recorded by a voltmeter.
We then notice that it is natural to separate the Hilbert space into
"inside-the-well" states and "outside-the-well"; to the latter we assign two states, "switched" ($m$) and "unswitched" ($\bar{m}$). In other words, for all our purposes in this paper, we will regard the states outside the well as constituting another qubit (an ancilla), whose interaction with the inside-the-well qubit is given by a controllable tunneling matrix element. For example, the state vector $|1\rangle|\bar{m}\rangle$ corresponds to the propagating wavepacket mentioned above. Finally, to keep the normalization, we write the evolution of the first excited state Eq. (\ref{pp}) under this measurement as
\begin{equation}
U|1\rangle|m\rangle = \sqrt{1-p}|1\rangle |m\rangle + \sqrt{p} |1\rangle |\bar{m}\rangle \label{unit}.
\end{equation}
Similarly, the state $|0\rangle |m\rangle$ corresponds to the qubit in the ground state and zero voltage across the junction, and it is left invariant under $U$,
\begin{equation}
U |0\rangle |m\rangle = |0\rangle |m\rangle .\label{unito}
\end{equation}
Formally, one can fully construct the operator $U$ under the unitarity condition; this results in
\begin{eqnarray}
U|1\rangle|\bar{m}\rangle &=& -\sqrt{p}|1\rangle |m\rangle + \sqrt{1-p} |1\rangle |\bar{m}\rangle , \\
U |0\rangle |\bar{m}\rangle &=& |0\rangle |\bar{m}\rangle,
\end{eqnarray}
or in matrix form, in the basis $\{|0m\rangle , |1m\rangle , |0\bar{m}\rangle , |1\bar{m}\rangle \}$ (in this order),
\begin{equation}
U =
\left( \begin{array}{cccc}
1 & 0 & 0 & 0\\
0 & \sqrt{1-p} & 0 & \sqrt{p} \\
0 & 0 & 1 & 0 \\
0 & -\sqrt{p} & 0 & \sqrt{1-p}
\end{array} \right)
\end{equation}

With this explicit construction of $U$ in place, we can now make one more step forward and prove  our claim.
Indeed, the action of the operator $U$ introduced by Eqs. (\ref{unit},\ref{unito}) can be written equivalently as (see also \cite{nielsen})
\begin{equation}
U|\psi\rangle|m\rangle = M_{m} |\psi\rangle |m\rangle + M_{\bar{m}} |\psi\rangle |\bar{m}\rangle , \label{uu}
\end{equation}
where $|\psi\rangle = \alpha|0\rangle + \beta |1\rangle$ is an arbitrary wavefunction, and
 $M_{m}$ and $M_{\bar{m}}$ have been introduced before Eqs. (\ref{em},\ref{embar}).
With this identification it follows immediately that a projective measurement on the ancilla, giving the result $m$, will collapse Eq. (\ref{uu}) to
$M_{m}|\psi_{i}\rangle$ (normalized as in Eq. (\ref{after})) and this will happen with probability $P(m|\psi)$ as in Eq. (\ref{duo}). This shows that a partial measurement can be understood as a sharp measurement on an ancilla that has interacted with the quantum system.

It is worth pointing out at this point that in real experiments the qubit is completely destroyed when tunneling occurs; thus, strictly speaking, from the second term of Eq. (\ref{unit})
we would be tempted to say that if the outcome $\bar{m}$ occurs, the qubit is left in the state $|1\rangle$, which is not the case in reality. This however will not matter at all in the following, since we will all the time postselect only the results $m$.

\paragraph{Partial measurements from a quantum-information perspective}

It is interesting to observe that in such measurements apparently nothing has happened, in the sense that there was no macroscopically recordable result or event (the voltage across the junction has remained zero). More precisely, there has been no exchange of energy between the qubit and the ancilla - had a quanta of energy been transferred, this would have "switched" the junction in the finite-voltage state. Despite this, we still have to adjust our information about the system (as encoded in the wavefunction). But how can it be that the lack of occurrence of a (macroscopic) event - which could have happend but didn't -  results in a change in our description of the system?  One has to remember that we are in fact actively interrogating the system by performing manipulations at the classical level: we adiabatically bring the system from the situation in which none of the states $|0\rangle$ and $|1\rangle$ can tunnel to the situation in which only $|1\rangle$ can tunnel, we keep it there for some time $\tau$, then reverse it to the original situation. It is no wonder that we can extract classical information about the system. Also, one notices that during  the measurement time $\tau$ the Hamiltonian of the system has been altered through a change in boundary conditions: due to the fact that the tail of the wavefunction in state $|1\rangle$ is  not negligible outside the well, the effective Hamiltonian has acquired a tunneling component. But a change in the Hamiltonian does not mean that some energy exchange has occurred: it only means that it could have occurred. In some sense, one can say that the separation between energy and information --  namely that in order to describe probabilistic systems one should have a correspondence between bits of information and entities which are well-separated (distinguishable) in energy (or, equivalently, mass) -- is an artefact of the classical description of the world. In contradistinction, in the quantum description, information and energy are inextricable constitutive parts of the mathematical and conceptual description.


\paragraph{Entropy of a partial measurement}
Since the results of a partial measurement are random, we can associate to them the standard binary entropy function. Suppose we have state
$|\psi\rangle = \alpha |0\rangle + \beta |1\rangle $. Then Eqs. (\ref{uno},\ref{duo}) give
\begin{eqnarray}
P(\bar{m}|\psi) &=& p|\beta|^{2}, \\
P(m|\psi) &=& 1-p|\beta|^{2},
\end{eqnarray}
The  entropy associated with this process is
\begin{eqnarray}
H &=& - P(m|\psi) \ln P(m|\psi) -  P(\bar{m}|\psi) \ln P(\bar{m}|\psi)  \\
 & & = - p|\beta|^{2} \ln (p|\beta|^{2})- (1-p|\beta|^{2}) \ln (1- p|\beta|^{2}) .
\end{eqnarray}
One can see that the entropy of a partial measurement is obtained simply by rescaling the probability associated with the state $|1\rangle$ by a factor $p$.
As a result, the maximum entropy is obtained at $|\beta | = 1/\sqrt{2p}$,  ${\it i.e.}$ at a value of $|\beta |$ smaller by a factor of $\sqrt{p}$ then the value
$1/\sqrt{2}$ characterizing the maximum entropy for sharp measurements along $z$.

\paragraph{Reversal of partial measurements}

We now discuss the idea of reversal of a partial measurement. Given that the outcome $m$ has been obtained, the evolution of the wavefunction can be described by the nonunitary transformation $M_{m}$. For the case $p\neq 1$, this transformation admits an inverse, which can be noticed immediately to be
\begin{equation}
M_{m}^{-1} = |0\rangle\langle 0|+ \frac{1}{\sqrt{1-p}}|1\rangle\langle 1| .
\end{equation}
A very useful observation is that this inversion can be achieved physically by a combination of two unitaries ($X$-gates\footnote{We use the symbol $X$ and $\sigma_x$ to denote the same operator, namely the $x$ Pauli matrix; the first notation is used extensively in modern quantum information; the second notation is more traditional.}), and another nonunitary transformation of the same type and strength as the one we want to reverse. With $X=|0\rangle\langle 1| + |1\rangle\langle 0|$, we have
\begin{equation}
M_{m}^{-1} = \frac{1}{\sqrt{1-p}}XM_{m}X.
\end{equation}
Note that the process of partial measurement reversal is probabilistic: assuming we start with a state $|\psi\rangle$, the probability for a successful measurement-reversal experiment, involving first a measurement $M_{m}$ and then the reversal $XM_{m}X$ can be calculated in the same way as in Eqs. (\ref{uno},\ref{duo}), yielding $\langle \psi|(XM_{m}XM_{m})^{\dag}XM_{m}XM_{m}|\psi\rangle = 1-p$, a probability which is independent on the initial state.
That this is indeed the case one can see by calculating the corresponding conditional probabilities at each step of the process. Suppose we start with
$|\psi\rangle = \alpha |0\rangle + \beta |1\rangle $. The probabilities after the first measurement are Eqs. (\ref{uno},\ref{duo}),
\begin{eqnarray}
P(\bar{m}|\psi) &=& p|\beta|^{2} ,\\
P(m|\psi) &=& 1-p|\beta|^{2}.
\end{eqnarray}
If the result $m$ is obtained, the wavefunction after the measurement is given by Eq. (\ref{after}),
\begin{equation}
|\psi_{m}\rangle = \frac{\alpha}{\sqrt{1-p|\beta|^2}}|0\rangle + \frac{\beta\sqrt{1-p}}{\sqrt{1-p|\beta|^2}}|1\rangle .
\end{equation}
The application of the operator $X$ then only transforms $|0\rangle\rightarrow |1\rangle$ and $|1\rangle\rightarrow |0\rangle
$; to the resulting wavefunction we apply a second measurement operator $M_{m}$.
Using now Eqs. (\ref{uno},\ref{duo}) for the wavefunction $X|\psi_{i}^{m}\rangle$, we obtain that the probability of obtaining again the result $m$ is
\begin{equation}\
\langle \psi_{i}^{m} |X M_{m}^{\dag} M_{m} X|\psi_{i}^{m}\rangle = (1-p)/(1-p|\beta|^{2}).
\end{equation}
By the multiplication rule for conditional probabilities, the final probability of success for the whole process is given by $\langle \psi_{i}^{m} |X M_{m}^{\dag} M_{m} X|\psi_{i}^{m}\rangle P(m|\psi)=1-p$, thus we obtain indeed the result claimed above.

\section{Superposition and entanglement - which  one is the central quantum mystery?}
\label{sup}

In his famous pedagogical exposition of interference experiments, Feynman described superposition as the "central" (or "only") quantum mystery \cite{feynman}. On the other hand, a long list of physicists starting with Schr\"odinger and Einstein were deeply bothered by entanglement. Which of the two is more mysterious or more fundamental might be for sure a question of taste. However, the construction given above for the operator $U$ allows us to show immediately that entanglement is a consequence of the superposition principle applied to the specific state structure provided by Hilbert spaces and tensorial products. Indeed, let us consider Eq. (\ref{unit}) and Eq. (\ref{unito}). The first is a clear statement of the superposition principle applied to a wavefunction which is allowed to tunnel between the wells; it is simply a restatement of
Eq. (\ref{pp}). But the superposition principle can be applied also to the two states $|0\rangle$ and $|1\rangle$: since $U$ is linear, when starting with a general $|\psi \rangle = \alpha |0\rangle + \beta |1\rangle $ we have
\begin{equation}
U|\psi\rangle |m\rangle = \alpha |0\rangle|m\rangle + \beta\sqrt{1-p}|1\rangle |m\rangle + \beta \sqrt{p}|1\rangle|\bar{m}\rangle .\label{enta}
\end{equation}
This state is in general entangled, with concurrence ${\cal C} = 2|\alpha\beta |\sqrt{p}$. Thus entanglement is obtained by applying consecutively, on two different subspaces, the superposition principle. For example, Eq. (\ref{enta}) becomes a Bell state (maximally entangled) for $\alpha = \beta = 1/\sqrt{2}$, and $p=1$.
This shows also that  by adding a chain of measurement apparatuses and treating them quantum-mechanically we cannot solve the measurement problem \cite{me} ({\it i.e.} we cannnot explain the collapse of the wavefunction by using quantum mechanics). Indeed, for $p=1$ the ancilla is performing the standard von Neumann measurement on the qubit: but instead of explaining so to say the physics of the process of projection, we ended up with a two-qubit wavefunction Eq. (\ref{enta}), which again has to be "collapsed" according to the same quantum-mechanical rules we would like to explain.

\section{Hidden variables}
\label{hid}

Do partial measurements tell us anything interesting about hidden variables?
A first observation is that there are so far no general theorems that rule out hidden variables for single spin-1/2 systems: Bell's theorem needs two spins, Kochen-Spekker a spin of at least 1, and further generalizations (GHZ, W, cluster states, {\it etc.}) require even larger Hilbert spaces.
There exists in fact a simple hidden-variable model for two-level systems, invented by Bell and simplified  by Mermin \cite{bell,mermin}. We review this model in Appendix \ref{ap2}. In the Bell-Mermin model, a spin-1/2 is described by two unit vectors in real space, $\vec{n}$ and $\vec{h}$. The first vector denotes what  we normally would call the quantum state of the system: that is, it embeds the information about the preparation procedure that the experimentalist has control of. Since any state in a two-dimensional Hilbert space can be regarded as the eigenvector of the spin along a direction $\vec{n}$, $\sigma_{\vec{n}}|\uparrow\rangle_{\vec{n}}=|\uparrow\rangle_{\vec{n}}$, it follows that specifying $\vec{n}$ is enough to fully account for the preparation procedure. The second unit vector, $\vec{h}$, represents a variable which is not under the control of the experimentalist. Naturally, this means that it will have a uniform statistical distribution on the unit sphere. Although in the following we will not employ explicitly this model, it is useful to keep it in mind.

We now return to the problem of hidden variables. We want here to strengthen our argument that the peculiar structure of Hilbert space, which allows for quantum entanglement and which cannot be reproduced by local realistic theories, is the reason why realistic hidden-variable theories fail. We want to show that it is not possible to construct a theory in which we separate the classical knowledge about the state from an unknown (hidden-variable) knowledge. As mentioned before, what happens for quantum objects is that classical information (which relies on bits assigned to separate entities) and quantum information are inextricably blended. Any attempt to separate them is bound to give contradictions with the experiment.

Precisely, let us assume that the qubit and the ancilla are each described by a set of two vectors in the form $(\vec{n}, \vec{h})$ and  $(\vec{n}', \vec{h}')$, the first of which is the state (the vector on the Bloch sphere representing the state), and the other one the hidden variable ($\vec{h}$ for the qubit and $\vec{h}'$ for the ancilla). For example, if  the qubit is in the state $|0\rangle$ then its corresponding $\vec{n}$ vector is the unit vector $\hat{\vec{z}}$ pointing to the North Pole of the Bloch sphere (and for the state $|1\rangle$ we have $-\hat{\vec{z}}$). Similarly, if the state of the ancilla is $\vec{m}$ the vector $\vec{n}'$ is $\hat{\vec{z}}'$, while to $\hat{\vec{m}}$ we associate $-\hat{\vec{z}}'$.

We also consider for simplicity $p=1$: the ancilla can be regarded as an apparatus performing a sharp measurement on the qubit, and, in turn, the experimentalist performs a sharp measurement on the ancilla. We then have two experimental results which we have to codify in the theory: if the qubit's initial state is $|0\rangle$, then the ancilla, after interaction, will always be in the state $|m\rangle$; if the qubit is in $|1\rangle$ the ancilla will always end up in the state $|\bar{m}\rangle$. So we have
\begin{eqnarray}
[(\hat{\vec{z}},\vec{h}); (\hat{\vec{z}}',\vec{h}')]&\rightarrow &[(\hat{\vec{z}}, \bullet); (\hat{\vec{z}}', \bullet)] ,\label{th1}\\
~[(-\hat{\vec{z}}
,\vec{h});
(\hat{\vec{z}}',\vec{h}')]
&\rightarrow &[(-\hat{\vec{z}},  \bullet ); (-\hat{\vec{z}}', \bullet )], \label{th2}
\end{eqnarray}
where, since we do not know the dependence of the new hidden variables of the qubit and ancilla on the initial ones, we mark them with a bullet.
Suppose now that we start with the state $|+\rangle$ for the qubit, meaning that we would like to apply the superposition principle, much like above, on the theory
Eq. (\ref{th1},\ref{th2}).
 Since we do not have a mathematical model for the mechanism of interaction, we write generically
\begin{equation}
[(\hat{\vec{x}},\vec{h}); (\hat{\vec{z}}',\vec{h}')]\rightarrow [(\hat{\vec{v}}, \bullet ); (\hat{\vec{v}}', \bullet )] .\label{labe}
\end{equation}
Here $\vec{v}$ and $\vec{v}'$ are vectors that we do not know, but they are yielded deterministically from the initial state $|+\rangle$\footnote{One can use the fact that the manipulations we  are doing are in the end adiabatic and argue that in fact $\hat{\vec{n}}=\hat{\vec{x}}$. See Appendix \ref{ap1} for a development of this argument.}. Remember that these vectors embody our classical knowledge about the manipulations we do in the lab, and therefore they do not depend on the hidden variable. Now, irrespective to what are the values of the vectors Eq. (\ref{labe}), and irrespective of the value of the hidden variables and their dependencies, the important thing is that Eq. (\ref{labe}) shows that it is possible to find a direction $\vec{v}'$ in space along which, when measuring the ancilla, we always find +1. However, this is not the case. Indeed, quantum mechanics describes the resulting state as a Bell state, $1/\sqrt{2}(|0\rangle|m\rangle + |1\rangle|\bar{m}\rangle)$, and on this state a measurement of the ancilla along any direction gives the results $\pm 1$ with equal probability.

To conclude this section, we note that the so-called Leggett-Garg inequality and the more recent work done to elucidate its connection to generalized weak measurements \cite{leggett-garg} put in evidence a related contradiction for a single qubit, namely between the assumption of realism and that of the non-invasiveness of  the measurements.

\section{The principle of complementarity}
\label{comp}

Suppose we have the following problem: a dice is given to an experimentalist and the task is to find if the dice is loaded. It is easy to solve this problem: one simply rolls the dice enough many times, records all the numbers, and singles out any statistically significant deviation from 1/6. But what if  one is given an unknown  wavefunction $|\psi \rangle = \alpha |0\rangle + \beta |1\rangle $? It is known that is not possible to determine the coefficients $\alpha$ and $\beta$ by performing several measurements on a single system, even if these measurements are weakly disturbing \cite{protective} ({\it e.g.} off-resonant homodyne measurements). The cumulative effect of the weak disturbance is always strong enough to forbid us to know the wavefunction with a reasonable degree of certainty. This is again a consequence of the fact that in quantum mechanics information is not embedded in distinguishable entities: indeed, despite the fact that the system has been prepared somehow in the state $|\psi\rangle$ (meaning that a preparation procedure has been followed), in the absence of classical communication it is not possible to extract from the object thus prepared what the procedure was.

The fact that one cannot determine the wavefunction of single quantum systems is intimately related to Bohr's principle of  complementarity. The connection is easy to see: if it were possible to perform at the same time measurements on conjugate variables on a given system, then one could use them to extract the complete information about the wavefunction.  Bohr's complementarity principle has its mathematical expression in the Heisenberg uncertainty relations. For example, for a spin 1/2 (a qubit) these relations lead to inequalities such as $\Delta_x\Delta_y \geq |\langle \sigma_z\rangle|$ and circular permutations of the indices $x,y,z$. Here $\Delta_{x,y}$ is the standard deviation associated with the observable $\sigma_{x,y}$, $\Delta_{x,y} =\sqrt{\langle\sigma_{x,y}^2\rangle - \langle\sigma_{x,y}\rangle^2}$,
and $\langle \sigma_z\rangle$ is the average of the observable $\sigma_z$. While the derivation of these relations is mathematically straightforward, their meaning is not. In the early days of quantum physics, during his  debates with Einstein, Bohr attempted to explain these relations by the unavoidable disturbance induced in a system when trying to measure one observable.  In modern times, it was realized that this interpretation is in fact not accurate, and that the underpinnings of the complementarity principle is not dynamical but kinematical, being related to the mathematical structure of the Hilbert space and to which-way information (see {\it e.g.}
\cite{zubairy} for a review of the most relevant modern experiments on complementarity). One can see in fact that the mathematical derivation of the uncertainty principle does not say anything about the measurements being performed one after the
other or simultaneously on a single quantum object: the standard deviation of the non-commuting observables such as $\sigma_x$ and $\sigma_y$ are calculated on different quantum objects of the same statistical ensemble, and not by performing consecutive measurements on the same quantum object \cite{kraus}.
The existence of unsharp measurements brings however a new twist to the problem:
if we regard these measurements as an extension of the standard quantum mechanical formalism, wouldn't then one hope to get more information about complementary observables than allowed by the uncertainty principle? Can one for example
get information about two complementary observables by measuring a third one?
The answer, unfortunately, is negative \cite{uffink}. Of course, if one accepts the loss of information due to the unsharpness of the measurements, the concept of joint measurements of complementary observables might still be useful in context such as quantum cryptography \cite{andersson}.

Here we suggest that new insights into the problem of complementarity can be gained by looking at what happens when we reverse a partial measurement.
We define the basis of the eigenvectors of the $\sigma_x$ Pauli operator by $\sigma_{x}|\pm\rangle = \pm |\pm\rangle$,
\begin{equation}
|+\rangle = \frac{1}{\sqrt{2}}\left( |1\rangle + |0\rangle\right),~~ |-\rangle = \frac{1}{\sqrt{2}}\left( |0\rangle - |1\rangle\right).
\end{equation}
Consider now the observables $\sigma_z$ and $\sigma_x$, and suppose we start with an initial state along the $x$ axis, say $|+\rangle = (1/\sqrt{2}) (|0\rangle + |1\rangle )$, which is an eigenvector of the operator $\sigma_x$ with eigenvalue $1$, $\sigma_{x}|+\rangle = |+\rangle$. After the measurement, the state becomes
\begin{equation}
|+\rangle \rightarrow \frac{1}{\sqrt{2-p}}\left(|0\rangle + \sqrt{1-p}|1\rangle\right) .
\end{equation}
What kind of information did we extract by performing this measurement? What we know now is that if we choose to measure $\sigma_z$ by a projective measurement, we will obtain the result "0" with probability $1/(2-p)$ and the result "1" with probability $(1-p)/(2-p)$. Note that $p$ could be in principle arbitrarily close to 1, meaning in this case that we have measured the value of $\sigma_z$ with arbitrarily good accuracy. Then, we reverse the measurement by the sequence $XM_{m}X$ as described above, and we keep only the qubits which have not "switched" ({\it i.e.} for which the result $m$ is again obtained. Now we are back to the state $|+\rangle$, and a measurement of $\sigma_x$ would confirm that we can measure $\sigma_x$ and obtain $+$. It now looks as if we have just contradicted the Heisenberg uncertainty principle. Is there a paradox here?

The reverse question is equally interesting: once we get the information that the spin is say in the state $|0\rangle$ (with good enough certainty), what happens with this information when we undo the measurement \cite{royer}? In this vein, we note that interference experiments can be understood in terms of {\it welcher-weg} (which-way) information. Acquiring information about whether the photon has passed through one slit or the other results in the destruction of the interference pattern. Here, it looks like we had it both ways: we did extract information about
which state the quanta has been into, and then we managed to fool the system into restoring its coherence as if nothing has been measured. The information about the qubit being in the state $|0\rangle$ has just vanished!

Next, we ask the question of weather partial measurements can help in determining the wavefunction of a single quantum object. As we have seen, this is not  possible with homodyne measurements. First, we notice that the effects in our scheme can be rewritten in terms of the identity and the $\sigma_z$ Pauli matrix:
\begin{eqnarray}
E_{m}&=&1 - \frac{p}{2}(1-\sigma_{z}) ,\\
E_{\bar{m}}&=&\frac{p}{2}(1-\sigma_{z}).
\end{eqnarray}
Then, a natural question to ask is: how to extend our scheme to partial measurements along an arbitrary direction? The answer is obvious: by performing appropriate rotations (single qubit gates) we can have the same form of the effects and measurement operators along any direction in space with the same partial measurement strength $p$. Suppose for example we want partial measurements along the $x$- axis. Then we simply apply a Hadamard gate,
\begin{equation}
H = \frac{1}{\sqrt{2}}
\left( \begin{array}{cc}
1 & 1\\
1 & -1
\end{array} \right),
\end{equation}
 to Eqs. (\ref{em}, \ref{embar}) and obtain the new
 measurement operators
\begin{eqnarray}
HM_{\bar{m}}H &=& \sqrt{p}|-\rangle\langle -|, \\
HM_{m}H &=& |+\rangle\langle +| + \sqrt{1-p}|-\rangle\langle -| .
\end{eqnarray}
Clearly, such measurements can also be reversed by first applying the inverse rotation and then undoing $M_{m}$ as before.

Now suppose that we start with an unknown state $\alpha |0\rangle + \beta|1\rangle$ and we perform a partial measurement. We extract the value of $\sigma_z$ and then undo the measurement. We can now rotate the state along any direction, do again a partial measurement, undo the measurement, and rotate back the state. It looks as if one can extract an arbitrary amount of information about the state, thus supporting the idea that the wavefunction is physical \cite{protective}.

Let us now now examine a bit closer this idea. Being successful in the procedure for reversal is a matter of chance: but one can imagine that we got lucky enough and, say, for the first qubit we have tried, we managed to do and undo all the measurements that we want without switching. But then what have we actually measured? How do we make use of the information we acquired? The answer is that, if one looks at the "successful" qubit only, there is in fact no information that can be extracted in this way. To understand why this is the case, let us examine the standard quantum tomography procedure for a single qubit. The procedure is as follows \cite{nielsen}: in general, for a mixed state described by a density matrix $\rho$, we use the expansion
\begin{equation}
\rho = \frac{1}{2}\left[Tr(\rho) + Tr(\rho \sigma_x) + Tr(\rho \sigma_y) + Tr(\rho \sigma_z)\right] ,
\end{equation}
and notice that $Tr(\rho \sigma_{x,y,z})$ are the average values of $\sigma_{x,y,z}$ measurements (while $Tr(\rho )$ can be obtained by summing over the probabilities
of getting the results 0 and 1 under a $\sigma_z$ measurement). For pure states, $|\psi \rangle = \cos(\theta /2)|0\rangle + \sin (\theta /2) \exp(i \varphi )|1\rangle$, and by the procedure above we have
\begin{eqnarray}
\cos \theta &=& \langle \sigma_z \rangle ,\\
\sin \theta \cos\varphi &=& \langle \sigma_x \rangle ,\\
-\sin \theta \sin\varphi &=& \langle \sigma_y \rangle .
\end{eqnarray}
The first two relations determine $\theta$ and $\varphi$ up to the sign of $\sin (\varphi )$, which is fixed by the measurements of $\sigma_y$ (third relation). Now, by using partial measurements it is also possible to do exact tomography.  We measure  first along $\sigma_z$, and get
\begin{equation}
\cos \theta = -1 + \frac{1 + P(\bar{m}|\psi) + P(m|\psi)}{p}, \label{urd}
\end{equation}
which allows us to determine $\theta$, and by doing before the measurement a Hadamard gate we get $\varphi$ up to the sign of $\sin\varphi$,
\begin{equation}
\sin \theta \cos \varphi = \cos\theta + \frac{P(m|H\psi) -P (\bar{m}|H\psi)-1}{p}. \label{urdd}
\end{equation}
"Unsharp" measurements do not imply that we cannot do precise quantum tomography!

But now, returning to our problem, instead of measuring say N qubits belonging to an ensemble, we take a single qubit, do a first measurement, reverse it, do a second measurement, reverse it, and so on. This is possible, with a certain (low) probability, and there is a chance that we stumble upon a qubit which luckily doesn't switch even after performing N times this procedure. Wouldn't this allow us to have performed full tomography on a single qubit?
The answer is that we simply cannot evaluate the probabilities entering in Eq. (\ref{urd}, \ref{urdd}) by the doing and undoing of the $N$ measurements on a single object. All we can say was that we were lucky enough to get the results $m$: but this is by no means an evaluation of any probability. The situation is equivalent to saying that, luckily, we got the same result for measurements on the first $N$ qubits we pick from an ensemble of qubits, each prepared in the state $|\psi\rangle$. In order to have any meaningful estimate of these probabilities, we have also to count the number of times in which we have failed to reverse the measurement. In order to extract probabilities, one should be able to count also the cases in which $\bar{m}$ occurs, and then undo the measurement. But this is not allowed by our measurement scheme.

\section{The EPR paradox}
\label{epr}

We consider two experimentalists, Alice and Bob, sharing two qubits entangled in a Bell state. Both Alice and Bob can perform measurements on their qubits; but Alice can perform partial measurements, undo the measurement, then do a projective measurement in the end. Bob is only doing projective measurement.

To characterize entanglement we use the concurrence \cite{concurrence}, defined for pure bipartite states $|\psi\rangle$ as  ${\cal C}(\psi ) = |\langle \psi|\sigma_{y}\otimes \sigma_{y}|\psi^{*}\rangle|$, where $|\psi^{*}\rangle$ is the complex conjugate of $|\psi\rangle$. The Bell basis is denoted by
\begin{eqnarray}
|\Phi^{\pm}\rangle = \frac{1}{\sqrt{2}}(|00\rangle \pm |11\rangle , \label{bell1}\\
|\Psi^{\pm}\rangle = \frac{1}{\sqrt{2}}(|01\rangle \pm |10\rangle ,\label{bell2}
\end{eqnarray}
What happens when we do a partial measurement for example on the first qubit?
We have
\begin{eqnarray}
|\Phi^{\pm}\rangle \rightarrow \frac{1}{\sqrt{2-p}}(|00\rangle \pm \sqrt{1-p}|11\rangle ,\\
|\Psi^{\pm}\rangle = \frac{1}{\sqrt{2-p}}(|01\rangle \pm \sqrt{1-p}|10\rangle .
\end{eqnarray}
For any of the Bell state, the concurrence changes in this process from 1 (maximum entanglement for Bell states) to
\begin{equation}
{\cal C}=\frac{2\sqrt{1-p}}{2-p} .
\end{equation}
Thus the degree of entanglement decreases, which is expected. One can also see this by forming Bell inequalities: suppose we look at the $|\Phi^{\pm}\rangle$ state: with the notation $\sigma_{\pm} = (\sigma_{x}\pm \sigma_{y})/\sqrt{2}$ we have
\begin{eqnarray}
\langle \Phi^{+}_{m}|\sigma_{x}\otimes \sigma_{\pm}|\Phi^{+}_{m}\rangle&=& \frac{\sqrt{2}}{2}\frac{2\sqrt{1-p}}{2-p} ,\\
\langle\Phi^{+}_{m} |\sigma_{y}\otimes \sigma_{\pm}|\Phi^{+}_{m}\rangle &=& \pm \frac{\sqrt{2}}{2}\frac{2\sqrt{1-p}}{2-p} ,
\end{eqnarray}
and we can built a CHSH inequality
\begin{equation}
|\langle \Phi^{+}_{m}|\sigma_{x}\otimes \sigma_{+} +
 \sigma_{x}\otimes \sigma_{-}
\sigma_{y}\otimes \sigma_{-}
-\sigma_{y}\otimes \sigma_{+}|\Phi^{+}_{m}\rangle|\geq 2 .
\end{equation}
The left hand side is $2\sqrt{2}{\cal C}$, therefore it decreases while ${\cal C}$ is decreasing.

This is not so surprising. Also a sharp measurement would decrese (to zero) the entanglement of two qubits. What is surprising however is what happens
 when we undo this measurement: the degree of entanglement increases back to 1, even if the reversal operation is local! We note here that there exists a related effect, that of creating entanglement between two qubits by performing conditional measurements on a third quantum object after they all have interacted in the past \cite{entanglementbymeasurement}, but perhaps in the case here the situation is more striking, since only two objects are involved and there is no interaction.
Also, we  notice that local unitary transformations do not change the degree of entanglement. Clearly, one could use  this effect to amplify a small degree of entanglement by performing partial measurements. What happens actually here? One can intuitively regard this procedure as a way to keep (by a type of post-selection) only the qubits which are highly entangled and throw away the others.

Now the EPR argument can be cast in the following form: a partial measurement on the first qubit by Alice will provide information (with a reasonable probability) about the value of $\sigma_z$ of Bob's qubit. For the state $|\Phi^{+}\rangle$ for example, one can predict the value of $\sigma_z$ of Bob's qubit to be 0 with probability $1/(2-p)\geq 1/2$, approaching  $1$ as $p$ gets large (close to $1$).
Now Alice reverses her measurement and after that measures $\sigma_x$ projectively, resulting in a definite value of Bob's qubit in a $\sigma_x$ measurement\footnote{One can argue here that there is still an element of counterfactual reasoning left: immediately after Alice's partial measurement Bob should in fact measure the state of his qubit and prove that it is close to  $|0\rangle$. This can be done on an ensemble (quantum tomography), but, as shown in this paper, not on a single qubit.}! Thus, during the entire procedure, Bob's qubit has been made to acquire (with reasonable certainty) two definite values, corresponding to non-commutating observables.

\section{Quantum teleportation}
\label{tel}

\begin{figure}[t]
\begin{center}
  \includegraphics[width=10cm]{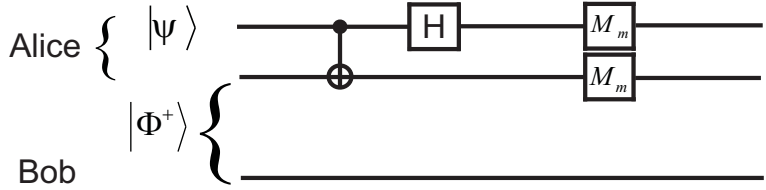}
\end{center}
\caption{Schematic of a teleportation experiment which uses partial measurements.}
         \label{fig2}
\end{figure}

In the standard version of teleportation \cite{nielsen}, Alice wants to teleport the quantum state $|\psi\rangle = \alpha |0\rangle + \beta |1\rangle$ using a second qubit which is
maximally entangled (in a $|\Phi^{+}\rangle$ state) with another qubit located somewhere far away, in
Bob's lab. By using a CNOT gate followed by a Hadamard gate, Alice creates the state
\begin{equation}
\frac{1}{2} \left[|00\rangle (\alpha |0\rangle + \beta |1\rangle ) + |01\rangle (\alpha |1\rangle + \beta |0\rangle ) + |10\rangle (\alpha |0\rangle - \beta |1\rangle ) + |11\rangle (\alpha |1\rangle - \beta |0\rangle )\right], \label{qqq}
\end{equation}
where the last qubit is Bob's qubit (see Fig. \ref{fig2}). Suppose now Alice performs two partial measurements on her two qubits with the result $00$; she communicates the result to Bob. Then she reverses the measurements, and does a further projective measurement on her qubits. This time, she obtains say the result $11$, which she is also communicating to Bob. How should Bob think about his qubit? If he trusts the first result, he would say that his qubit is in  $\alpha|0\rangle + \beta |1\rangle$. If he trusts the second, it is $\alpha|1\rangle - \beta |0\rangle$. Like in the EPR analysis, it looks as if Alice controls the reality of Bob's qubit remotely.

Moreover, it looks at first sight that one has obtained a cloning machine: if Alice does partial measurements with large strengths, Bob's qubit will be, to a good degree of approximation, in  the same states as for the standard case of sharp measurements. But now Alice reverses her measurements and, since the H gate and the CNOT gates are reversible, she can apply them again and thus
she recovers her initial state. This of course is not so for the cases in which Alice  fails to undo her measurements, but the point is that sometimes it happens, and we end up with a quite simple cloning machine.

However, this is not the case. Let us look at what happens when Alice applies two partial measurements, of strengths $p$ and $\tilde{p}$, on each of her qubits. The state Eq. (\ref{qqq}) transforms into
\begin{eqnarray}
& & \frac{1}{2} \left[|00\rangle (\alpha |0\rangle + \beta |1\rangle ) + \sqrt{1-\tilde{p}}|01\rangle (\alpha |1\rangle + \beta |0\rangle ) +\right.\nonumber \\
 & & \left.
+ \sqrt{1-p}|10\rangle (\alpha |0\rangle - \beta |1\rangle ) + \sqrt{(1-p)(1-\tilde{p})}|11\rangle (\alpha |1\rangle - \beta |0\rangle )\right] \label{smile}
\end{eqnarray}
By inspecting  this state, one can say that, if $p$ and $\tilde{p}$ are large enough, Bob's qubit can be arbitrarily close to the state $|\psi\rangle$. Does that mean that, for all practical purposes, the qubit is in this state? Let us see now what happens when we reverse the measurement: we apply, as usual, the combinations $XM_{m}X$ at Alice's qubit, with strengths $p$ and $\tilde{p}$. It is immediate to check that we get back the state Eq. (\ref{qqq}) and one can indeed apply again the Hadamard gate and the CNOT gate to recover Alice's initial state. One can see however that during the "undoing" process Bob's state has changed also: it  is no longer arbitrarily close to $|\psi\rangle$ but now it is back as part of a $|\Phi^{+}\rangle$ Bell state! Unfortunately, the state reversal performed by Alice has reconstructed not only the unknown state of her first qubit,
but the state of the whole three-qubit system, even if Bob did not do anything. It is somewhat surprising that by acting only on her two qubits in a distant place, Alice has managed to reconstruct the wavefunction of a system which can be extremely delocalized!
What is happening here is that the small correlations due to the terms $|01\rangle , |10\rangle , |11\rangle $ in Eq. (\ref{smile}) became amplified by the process of "undoing". One might say that exactly those qubits that have managed to survive through the reconstruction process are the ones which give the "errors" in approximating Eq. (\ref{smile}) with $|00\rangle (\alpha |0\rangle + \beta |1\rangle )$.

But  what if now Bob is allowed to do a sharp measurement of his qubit? Does it matter if he is doing it before  or after Alice's measurement reversal? If Bob measures before Alice, will Alice still be able to reverse the "collapsed" wavefunction resulting from Bob's impetuous behavior? Suppose that after Alice's successful reversal of measurement, Bob measures the state $|0\rangle$. Then the state of Alice's qubits is, according to
Eq. (\ref{qqq}),
\begin{equation}
\alpha |+\rangle |0\rangle + \beta |-\rangle |1\rangle . \label{final}
\end{equation}
Suppose now that Bob measures the same state after Alice's partial measurements but before she manages to undo them. The (unnormalized) collapsed state resulting from Eq. (\ref{smile}) is then
\begin{equation}
\alpha |00\rangle + \sqrt{1-\tilde{p}}\beta |01\rangle
+ \sqrt{1-p}\alpha|10\rangle - \sqrt{(1-p)(1-\tilde{p})}\beta |11\rangle .\label{collapsed}
\end{equation}
Now Alice applies the combinations $XM_{m}X$ in order to undo the measurements. It is immediate to check that the state Eq. (\ref{collapsed}) is transformed into Eq. (\ref{final}). Thus, the temporary order of Alice's and Bob's measurements is irrelevant. Moreover, if now Alice passes the state Eq. (\ref{final}) backwards through a Hadamard gate and a CNOT, she gets:
\begin{eqnarray}
&& \alpha |+\rangle |0\rangle + \beta |-\rangle |1\rangle \\ && \stackrel{\rm H}{\rightarrow}
\alpha |0 |0\rangle + \beta |1\rangle |1\rangle \\
&& \stackrel{\rm CNOT}{\longrightarrow}
(\alpha|0\rangle + \beta |1\rangle)|0\rangle .
\end{eqnarray}
So she did get her qubit back, but note that Bob has not gotten anything out of  this story: his qubit is in the state $|0\rangle$, and the result would have been the same had he measured his qubit at the very beginning of the experiment, when the state of the 3-qubit system was $|\psi\rangle|\Phi^{+}\rangle$.

\section{Entanglement swapping}
\label{swap}

Entanglement swapping was first proposed in \cite{zukowski} and later demonstrated experimentally \cite{pan}. It is a procedure in which two particles, each maximally entangled with a EPR partner, become entangled {\it via} measurements on their partners.

\begin{figure}[t]
\begin{center}
  \includegraphics[width=10cm]{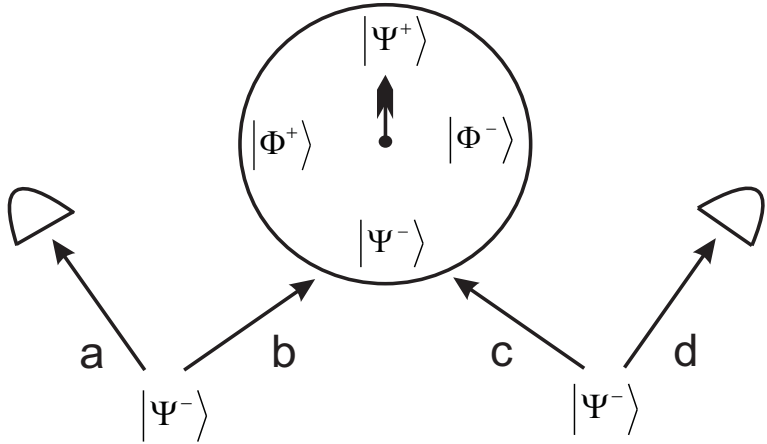}
\end{center}
\caption{Entanglement swapping: particles $b$ and $c$ enter a Bell measurement setup (represented here as a clock with four readings) while particles $a$ and $d$ are detected separately.}
         \label{fig3}
\end{figure}

Suppose we denote the Bell basis as before by $|\Phi^{\pm}\rangle$, $|\Psi^{\pm}\rangle$ as in Eq. (\ref{bell1},\ref{bell2}), and we index the four particles by $a$, $b$, $c$, and $d$. Suppose that $a-b$ are maximally entangled in a Bell singlet $|\Psi^{-}\rangle_{ab}$ and that $c-d$ are as well in a Bell singlet state
$|\Psi^{-}\rangle_{cd}$. The state to be measured is therefore $|\Psi^{-}\rangle_{ab}|\Psi^{-}\rangle_{cd}$, clearly showing that the particles $a-d$ for example have  not seen each other. But this state can be expanded as
\begin{equation}
|\Psi^{-}\rangle_{ab}|\Psi^{-}\rangle_{cd} = |\Psi^{+}\rangle_{ad}|\Psi^{+}\rangle_{bc} - |\Psi^{-}\rangle_{ad}|\Psi^{-}\rangle_{bc} -
|\Phi^{+}\rangle_{ad}|\Phi^{+}\rangle_{bc} +
|\Phi^{-}\rangle_{ad}|\Phi^{-}\rangle_{bc} . \label{initial}
\end{equation}
If one could have a measurement setup which projects onto the Bell basis (something which can be done with quantum-optics techniques), then by measuring particles $b-c$ and getting for example the result $|\Psi^{+}\rangle_{bc}$ we have projected
the initial state Eq. (\ref{initial}) to $|\Psi^{+}\rangle_{ad}$, getting $a-d$ into an entangled state. A schematic of this experiment is shown in Fig. \ref{fig3}. This clearly violates our intuition that particles must interact in order to get entangled.

There is one more element of surprise in this story: the measurement of the particles $b-c$ can be taken anytime - even after $a-d$ have been recorded and they might not even exist (if for example they are photons). One can also delay the decision of weather to measure them in the Bell basis or leave them in the original form: in general, the degree to which the particles are entangled can be defined after they are detected \cite{delay}. By performing the two sets of measurements under the locality condition, one ensures that there is no hidden causal connection between the act of measuring and the result of obtaining entanglement \cite{ma}.

With the help of partial measurements we add one more layer of paradox to this: one immediately notices that, by using partial measurement and for a given set of four particles, we can get them entangled and then, if the experimentalist has a change of mind,  him/her can disentangle them back.
The explanation of all the above is that we cannot conclude anything from a single set of four measurements. The entangled states  $|\Psi^{\pm}\rangle_{ad}$ $|\Phi^{\pm}\rangle_{ad}$ tell us that Bell correlations are obtained by partitioning the total ensemble into subsets which are in correspondence with the results $|\Psi^{\pm }\rangle_{bc}$, $|\Phi^{\pm}\rangle_{bc}$. This partitioning can be done in different ways, depending on the information coming from the experimentalist who has measured $b -c$.

\section{Conclusions}
\label{conc}

Partial measurements and their reversal are an interesting tool for exploring the foundations of quantum physics. In this paper we highlight a series of intuitively puzzling results that can be obtained when replacing von Neumann measurements with partial measurements in standard quantum-physics experiments. As always with quantum physics, there is no real logical paradox: in most of the situations we describe, the crux of  the problem is our need to think about the wavefunction as something real, existing also in single quantum objects. But the image of a particle that "carries" its own wavefunction (or the other way around) is incorrect. The proper way to think about the wavefunction seems to be as just a mathematical tool to characterize the probabilities of obtaining a result for an ensemble of particles.

\section{Acknowledgements}

 The research for this work has started under a John Templeton Fellowship, which allowed the author to spend the summer of 2009 at IQOQI Vienna. Special thanks go to my
hosts in Vienna, Prof. A. Zeilinger and Prof. M. Aspelmeyer, who have
made this visit possible, and additionally to the scientists in the institute for many enlightening discussions. I also thank Raymond Chiao and Alexei Grinbaum for useful discussions. Also, support from the Academy of Finland is acknowledged (Acad. Res. Fellowship 00857, and projects 129896, 118122, and 135135).

\appendix

\section{Generalized partial measurements}
\label{ap1}

A natural generalization of partial measurements as introduced in Section \ref{part} is by allowing also the states $|0\rangle$ to tunnel out with a certain probability $q$.
This can be realized experimentally by raising the bias current in the junction to a value such that the corresponding potential formed allows tunneling of both states.
Formally, we will have instead of Eq. (\ref{em},\ref{embar}),
\begin{eqnarray}
M_{\bar{m}} (p,q)&=& \sqrt{q}|0\rangle\langle 0| + \sqrt{p}|1\rangle\langle 1| , \label{mg}\\
M_{m} (p,q) &=& \sqrt{1-q}|0\rangle\langle 0| + \sqrt{1-p}|1\rangle\langle 1| , \label{embarg}
\end{eqnarray}
corresponding to the effects
\begin{eqnarray}
E_{\bar{m}} (p,q)&=& M_{\bar{m}}^{\dag}M_{\bar{m}} =  q |0\rangle\langle 0|+p |1\rangle\langle 1|,   \\
E_{m} (p,q)&=& M_{m}^{\dag}M_{m} = (1-q) |0\rangle\langle 0| + (1-p) |1\rangle\langle 1|,
\end{eqnarray}
 which again provide a semispectral resolution of the identity
 $E_{\bar{m}}(p,q) +E_{m}(p,q)=1$.
 For  a state $|\psi\rangle = \alpha |0\rangle + \beta |1\rangle$ we
 obtain the probabilities
\begin{eqnarray}
P(\bar{m}|\psi) &=& \langle \psi |E_{\bar{m}}(p,q)|\psi\rangle = |\alpha|^2 q + |\beta|^2 p ,
\label{unog}\\
P(m|\psi) &=& \langle \psi |E_{m}(p,q)|\psi\rangle =1- (|\alpha|^2 q + |\beta|^2 p) ,
\label{duog}
\end{eqnarray}
and the wavefunctions after the measurement
\begin{eqnarray}
|\psi_{\bar{m}}\rangle &=& \frac{1}{\sqrt{P(\bar{m}|\psi)}}M_{\bar{m}}(p,q)|\psi\rangle =\frac{1}{\sqrt{P(\bar{m}|\psi)}}\left[\alpha \sqrt{q}|0\rangle + \beta \sqrt{p}|1\rangle \right],
\label{aftershaveg}\\
|\psi_{m}\rangle &=& \frac{1}{\sqrt{P(m|\psi)}}M_{m}(p,q)|\psi\rangle
= \frac{1}{\sqrt{P(m|\psi)}}\left[\alpha \sqrt{1-q}|0\rangle + \beta \sqrt{1-p}|1\rangle\right] .
 \label{afterg}
\end{eqnarray}
In a way similar to the discussion in Section (\ref{part}) we can introduce an ancilla and show the equivalence of these generalized measurements with evolution on the extended Hilbert space followed by sharp measurements on the ancilla. The corresponding operator $U$ is now
\begin{equation}
U (p,q)=
\left( \begin{array}{cccc}
\sqrt{1-q} & 0 & \sqrt{q} & 0\\
0 & \sqrt{1-p} & 0 & \sqrt{p} \\
-\sqrt{q} & 0 & \sqrt{1-q} & 0 \\
0 & -\sqrt{p} & 0 & \sqrt{1-p}
\end{array} \right).
\end{equation}

Generalized partial measurement can be as well inverted. This time however, for $p, q\neq 0,1$ the reversal of the measurement can be done also if the result $\bar{m}$ is obtained. Explicitly, we have
\begin{eqnarray}
M_{m}^{-1}(p,q) &=& \frac{1}{\sqrt{(1-p)(1-q)}}XM_{m}(p,q)X , \\
M_{\bar{m}}^{-1}(p,q) &=& \frac{1}{\sqrt{pq}}XM_{\bar{m}}(p,q)X .
\end{eqnarray}

Generalized partial measurements can be obtained by a sequence of partial measurements of  the type presented in Eq. (\ref{em},\ref{embar}). For clarity, we will here the operators Eq. (\ref{em},\ref{embar}) as a function of the strength $p$: $M_{m}(p)$;  $M_{\bar {m}}(p)$, {\it etc.}.
With these notations, we have
\begin{equation}
M_{m}(p,q) = XM_{m}(p)XM_{m}(q), \label{rt}
\end{equation}
which physically means that, when  applying two $M_{m}$ operations and two X gates in the order shown in Eq. (\ref{rt}) (or, equivalently, as $M_{m}(q)XM_{m}(p)X$), we obtain a  $M_{m}(p,q)$ operation conditional on the qubit not switching. Note that $M_{\bar{m}}(p,q)$ is not obtained as
$XM_{\bar{m}}(p)XM_{\bar{m}}(q)$, but by
\begin{equation}
M_{\bar{m}}(p,q) = XM_{\bar{m}}(p)XM_{m}(q) + XM_{m}(p)XM_{\bar{m}}(q)+ XM_{\bar{m}}(p)XM_{\bar{m}}(q),
\end{equation}
showing that a final switching state is obtained either by a switch in the first or in the second partial measurement.

\section{The Bell-Mermin model for spin-1/2 hidden variables and partial measurements}
\label{ap2}

John Bell \cite{bell} was the first to realize that it is perfectly possible to invent a hidden-variable model for a single spin-1/2; an elegant version of this model has been put forward by David Mermin \cite{mermin}. Here we give a brief review of this model and show a particular form of the argument presented in Section \ref{part}. A spin-1/2 observable $A$ can be written as $A=a_{0} + \vec{a}_{1}\vec{\sigma}$ ($a_{0}$ is a scalar and $\vec{a}_1$ is a vector of magnitude $|\vec  {a}_{1}|$), and its measured values are  $v(A) = a_{0}\pm |\vec  {a}_{1}|$. The average of $A$ on a state $|\uparrow\rangle_{\vec{n}}$ (an eigenstate of the spin along the direction $\vec{n}$) is given, according to quantum mechanics, by $\langle A\rangle_{\vec{n}}\equiv~_{\vec{n}}\langle\uparrow|A|\uparrow\rangle_{\vec{n}} = a_{0} + \vec{a}_{1}\vec{n}$. In order to account for these experimental results, the Bell-Mermin model postulates a "measurement theory" which uses the hidden variable $\vec{h}$, namely
\begin{equation}
v_{\vec{n}}(A) = \left\{\begin{array}{cc} a_{0} + |\vec{a}_{1}|,  & (\vec{n}+\vec{h})\vec{a}_{1}>0 ,\\ a_{0} - |\vec{a}_{1}|,  & (\vec{n}+\vec{h})\vec{a}_{1}<0 . \end{array}\right.
\end{equation}
With this, one obvioulsy gets the same individual results $a_{0} \pm |\vec{a}_{1}|$ as observed experimentally, and by averaging
over the solid angle $\Omega_{\vec{h}}$ (see e.g. Appendix A in \cite{ff}) one obtains an identical value to the one given by quantum mechanics,
$
\langle A\rangle_{\vec{n}} = \int (d\Omega_{\vec{h}}/4\pi)v_{\vec{n}}(A)
=a_{0}+\vec{a}_{1}\vec{n}.
$

We now apply this model to our partial measurements. In this case, $A=\sigma_{z}$, therefore $a_{0}=0$, $\vec{a}_{1}=\hat{\vec{z}}$, the unit vector pointing in the positive value of the $z$-axis.
To make the situation more clear, we consider the case of a relatively large $p$ (but still $p\leq 1$). In this case, under a partial measurement the resulting state will be approximately equal to $|0\rangle$. Now, when thinking in terms of the Bell-Mermin model one sees first that due  to the fact that the manipulations are adiabatic the vector
$\vec{n}$ must remain unchanged (remember that $\vec{n}$ embeds knowledge that  the experimentalist has access to: but adiabatic manipulations do not change the state and no other classical event has happened since the junction did not switch). What must have happened is that this procedure has selected the qubits with  $(\vec{h}+\vec{n})\hat{\vec{z}}\geq 0$ (indeed those with $(\vec{h}+\vec{n})\hat{\vec{z}}\leq 0$ must have switched, according to the measurement theory).  Suppose now that $\vec{n}=\hat{\vec{x}}$, {\it i.e.} the
system was initially prepared in the state $|+\rangle = (1/\sqrt{2})\left(|0\rangle + |1\rangle \right)$. This means  we have selected the qubits with $\vec{h}\geq 0$, {\it i.e.} $\vec{h}$ takes values only in the upper hemisphere of unit radius.
But this cannot be true for two reasons: the first is that half of the initial qubits have this property, while the proportion of qubits we have selected is much smaller (due to $p\approx 1$). The second is that if, after the selection we do a measurement of $\sigma_x$, according to the measurement theory the result should always be +1; indeed, $\hat{\vec{x}}(\hat{\vec{x}} + \vec{h}) = 1 + \hat{\vec{x}}\vec{h}\geq 0$. But on the other hand, according to quantum theory, the state after the measurement is close to $|0\rangle$, therefore negative and positive values for $\sigma_x$ measurements are equally probable.

\end{document}